\documentclass{article}
\usepackage{spconf,amsmath,graphicx}
\usepackage[ngerman]{babel}
\usepackage{amsmath}
\usepackage{amsfonts}
\usepackage{amssymb}

\def\N{{\mathcal{N}}}

\newcommand{\tildew}{}
\newcommand{\GSUM}[4]{\sum_{#1=0}^{#2} #3_{#1}\varphi_{#1} #4}
\newcommand{\GSUMM}[4]{\sum\nolimits_{#1=0}^{#2} #3_{#1}\varphi_{#1} #4}
\newcommand{\GSUMtilde}[4]{\sum_{#1=0}^{#2} #3_{#1}{\varphi}_{#1} #4}
\title{Multi-modal filtering for non-linear estimation}
\name{Sanket Kamthe$^1$, Jan Peters$^{1,2}$, and Marc Peter
  Deisenroth$^{3,1}$\thanks{The research leading to these results has
    received funding from the EC's Seventh Framework Programme under
    grant agreement \#270327.}} \address{
\begin{tabular}{ccc}
$^1$Department of Computer Science & $^2$MPI for Intelligent Systems &
$^3$Department of Computing\\
Technische Universit\"at Darmstadt & T\"ubingen & Imperial College
London\\
Germany & Germany & United Kingdom
\end{tabular}
}

\begin{document}
%
\maketitle
\begin{abstract}
  Multi-modal densities appear frequently in time series and practical
  applications. However, they cannot be represented by common state
  estimators, such as the Extended Kalman Filter (EKF) and the
  Unscented Kalman Filter (UKF), which additionally suffer from the
  fact that uncertainty is often not captured sufficiently well, which
  can result in incoherent and divergent tracking performance. In this
  paper, we address these issues by devising a non-linear filtering
  algorithm where densities are represented by Gaussian mixture
  models, whose parameters are estimated in closed form. The resulting
  method exhibits a superior performance on typical benchmarks.
%
%
\end{abstract}
\section{INTRODUCTION AND RELATED WORK}
\label{sec:intro}

Time series models, which infer latent space variables from noisy
observations, have been extensively studied. For linear estimation in
stationary and non-stationary time series models the Kalman
filter~\cite{kailath2000linear} has been shown to be highly efficient,
theoretically and practically. The Kalman filter is optimal for linear
Gaussian systems~\cite{Anderson2005}. In such systems, the
Gaussianity allows us to derive the recursive filtering equations in
closed-form.
In contrast, for a non-linear system Gaussian uncertainties may become
non-Gaussian due to the non-linear transform. Hence, we require
approximations, such as linearising the functions, e.g. in the
Extended Kalman Filter (EKF), or deterministic sampling, e.g. in the
Unscented Kalman Filter (UKF) \cite{julier2004unscented} to
approximate a non-Gaussian density by a Gaussian. Such approximations
make the limiting implicit assumption that the true densities are
uni-modal. Filters based on these approximations often severely
under-perform when true densities are multi-modal. Hence, multi-modal
approaches are frequently needed.

For representing multi-modal, non-Gaussian densities particle filters
\cite{doucet2001sequential} are a standard approach. They are
computationally demanding since they often require a large number of
particles for good performance, e.g. due to the curse of
dimensionality. An insufficient number of particles may fail to
capture the tails of the density and lead to degenerate solutions. In
practice, we have to compromise between the deterministic and fast
(UKF\slash EKF) or the computationally demanding and more accurate
Monte Carlo methods \cite{doucet2001sequential}.

An ideal filter for a non-linear system should allow for multi-modal
approximations, and at the same time its approximations should be
consistent to avoid degenerate solutions. In this paper, we propose a
filtering method that approximates a non-Gaussian density by a
Gaussian mixture model (GMM). Such a GMM allows modeling
multi-modality as well as representing any density with arbitrary
accuracy given a sufficiently large number of Gaussians,
see~\cite{Anderson2005}, Section 8.4, for a proof. The GMM
 presents an elegant deterministic filtering solution in
the form of the Gaussian Sum filter \cite{alspach1972nonlinear}.

The Gaussian Sum filter (GSUM-F) was proposed as a solution to
estimation problems with non-Gaussian noise or prior densities. The
GSUM-F relies on linear dynamics and the assumption that the
parameters of the Gaussian mixture approximation to the non-Gaussian
noise or prior densities are known a priori. This linearity assumption
can be relaxed, e.g.  by linearisation (EKF
GSUM-F)~\cite{Anderson2005} or deterministic sampling (UKF
GSUM-F)~\cite{luo2010scaled}, but both solutions still require a
priori knowledge of the GMM parameters. If, however, the prior and
noise densities are Gaussian, the UKF GSUM-F and EKF GSUM-F equal the
standard UKF and EKF, i.e. they become uni-modal filters. To account
for a possible uni-modal to multi-modal transition in a non-linear
system, we need to solve two problems: the propagation of the
uncertainty and the parameter estimation of the GMM
approximation. Kotecha and Duric~\cite{kotecha2003gaussian}, proposed
random sampling for uncertainty propagation and
Expectation-Maximization (EM) to estimate the GMM parameters. In this
paper, we propose to propagate uncertainty deterministically using the
Unscented Transform, which also allows for a closed-form expression of
the GMM parameters.

The main contributions of this paper are the derivation of the
Multi-Modal-Filter (M-MF), a multi-modal approach to filtering in
non-linear dynamical systems, where all densities are represented by
Gaussian mixtures. Moreover, we present closed-form expressions for
estimating the parameters of the Gaussian mixture
model. 


%
\section{SYSTEM MODEL}
\label{sec:system_model}
We consider nonlinear dynamical systems 
\begin{align}
\label{equ:sysfunc} x_{n} &= f (x_{n-1} ) + w_{n},\quad &w_{n} \sim
\mathcal{N}(0,Q), \\ 
y_{n} &= h (x_{n} ) + v_{n}, &v_{n} \sim \mathcal{N}(0,R),
\end{align}
where $f$ and $h$ are the non-linear transition and measurement
function, respectively. The noise processes $w_{n}$ and $v_{n}$ are
i.i.d. zero mean Gaussian with covariances $Q$ and $R$,
respectively. We denote the $D$ dimensional state by $x_{n}$, and
$y_{n}$ is the $E$ dimensional observation. $Y_{j}=\{y_{1},\ldots,
y_{j}\} $ represents all observations up to time step $j$. We define
the state estimation problem as determining the density
$p(x_{n}|Y_{j})$. Filtering and prediction are defined for $j=n$ and
$j<n$, respectively. We define a GMM representation of a state
distributions as
\begin{align}
  p(x_{n}|Y_{j})&:=
  \GSUMM{i}{M}{\delta}{(x_{n|j})}\,,\\
  \varphi_{i} (x_{n|j})&:=\N (x_{n|j} | \mu^{i}_{n|j},\Sigma^{i}_{n|j})\,,
\label{eq:stateGMM}
\end{align}
with weights $\delta_{i} \in
[0,1]\: \text{ and requiring }\sum_i{\delta_{i}}=1$, which ensures that
$p(x_{n}|Y_{j})$ is a valid probability distribution.

%
%
%
\section{MULTI-MODAL FILTERING}
\label{sec:Multimodal}
In the following, we devise a closed-form filtering algorithm with
multi-modal representations of the state distributions.  Our algorithm
is inspired by the following observation made by Julier and Uhlmann~\cite{julier2004unscented}: ``Given only the mean and the variance
of the underlying distribution, and, in absence of any a priori
information, any distribution (with the same mean and variance) used
to calculate the transformed mean and variance is trivially optimal.''
This observation was the basis to derive the Unscented transform and
the UKF. However, predictions based on the Unscented Transform often
under-estimate the true predictive uncertainty, which can result in
incoherent state estimation and divergent tracking performance. 

To address this issue, we use a different (optimal) representation of
the underlying distribution, which still matches the mean and
variance: We propose to represent each sigma point in the Unscented
transform by a Gaussian centred at this sigma point. This
approximation of the original distribution is effectively a GMM with
$2D+1$ components.

In Section~\ref{subsec:Parameter}, we derive an optimal GMM
representation of a state distribution $p(x_{n-1})$ of which only the
mean and variance are known.  In Section~\ref{subsec:Uncertainty}, we
detail how to map this GMM through a non-linear function to obtain a
predictive distribution $p(x_n)$, which is represented by a GMM.  We
generalise both uncertainty propagation and parameter estimation to
the case where $p(x_{n-1})$ is given by a GMM. In
Section~\ref{subsec:mxtrred}, we propose a method for pruning the
number of mixture components in a GMM to avoid their exponential
increase in number of components. In Section~\ref{sec:Gsum}, we
propose the resulting filtering algorithm, which exploits the results
from Sections~\ref{subsec:Parameter}--\ref{subsec:mxtrred}.

\subsection{Estimation of the Gaussian Mixture Parameters}
\label{subsec:Parameter}
Let the mean and the variance of the state distribution $p(x_{n-1})$
be given by $\mu$, $\Sigma$, respectively.  Then, we can represent
$p(x_{n-1})$ by a GMM
$\tildew{p}(x_{n-1})=\GSUM{i}{2D}{\delta}{(x_{n-1})}$, such that the
mean and the variance of the approximate density $\tildew{p}(x_{n-1})$
equal the mean $\mu$ and variance $\Sigma$ of $p(x_{n-1})$. This
representation is achieved by the closed-form relations
\begin{equation}
\label{equ:splitGauss}
\begin{aligned}
\delta_i &= 1/(2D+1), \\
\mu^{0}&=\mu, \quad
\mu^{j}=\mu + {\sigma}^{j},\quad \mu^{j+D} = \mu-\sigma^j,\\
\Sigma^i &= \left(1- \tfrac{ 2 \alpha}{D+1}\right)\Sigma,
\end{aligned}
\end{equation}
where $i = 0,\dotsc, 2D$ and $j = 1,\dotsc, D$, where $D$ is the
dimensionality of the state variable $x_{n-1}$. The variable
${\sigma}$ denotes $D$ rows or columns from the matrix square root
$\pm \sqrt{\alpha\Sigma}$. From \eqref{equ:splitGauss}, we can see
that we need to calculate $\sqrt{\Sigma}$ only once for all $2D+1$
Gaussians $\varphi_i(x_{n-1})$. To ensure that $\Sigma^{i}$ is
positive semi-definite, the scaling factor $\alpha$ should be chosen
such that $2\alpha\leq \left(2D+1\right)$, see~\eqref{equ:splitGauss}.
For $2\alpha=2D+1$ in \eqref{equ:splitGauss}, the equations above
reduce to scaled sigma points. Hence, the GMM representation
in~\eqref{equ:splitGauss} can be considered a generalisation of the
classical sigma point representation of densities employed by the
Unscented Transform, where each sigma point becomes an improper
probability distribution.

%
\subsection{Propagation of Uncertainty}
\label{subsec:Uncertainty}
A key step in filtering is the uncertainty propagation step,
i.e. estimating the probability distribution of random variable, which
has been transformed by means of the transition function $f$. Given
$p(x_{n-1})$ and the system dynamics~\eqref{equ:sysfunc}, we determine
$p(x_{n})$ by evaluating $\int p( x_n | x_{n-1} )p(x_{n-1}) \mathrm{d}
x_{n-1}$. For non-linear functions $f$, the integral above can only
rarely be solved in closed form. Thus, approximate solutions are
required.

Uncertainty propagation in non-linear systems can be achieved by
approximate methods, employing linearisation or deterministic sampling
as in the EKF and UKF. In such approaches, the state distribution
$p(x_{n-1})$ and the approximate predictive density
$\tildew{p}(x_{n})$ are well represented by Gaussians. If the state
distribution $p(x_{n-1})$ is a Gaussian mixture as
in~\eqref{equ:splitGauss}, we can estimate the predictive distribution
$p(x_n)$ similarly, e.g.  by applying such an approximate update to
each mixture component in the GMM, see~\eqref{eq:stateGMM}.  In this
paper, we propagate each mixture component $\varphi_{i}(x_{n-1})$ of
the GMM through $f$ and approximate $p(x_{n})$ by
\begin{equation}
\begin{aligned}
\label{equ:UTGauss}
p(x_{n})&= \int p(x_n|x_{n-1} )\GSUMM{i}{2D}{\delta}{(x_{n-1})}
\mathrm{d} x_{n-1} \\
&\approx \GSUMM{j}{2D}{\delta}{(x_{n})},
\end{aligned}
\end{equation}
where the mean and covariance of each $\varphi_j(x_n)$ are computed by
means of the Unscented Transform. 

If the prior density is a Gaussian mixture
$p(x_{n-1})=\GSUM{j}{M-1}{\beta}{(x_{n-1})}$, we repeat the procedure
above for each mixture component in $p(x_{n-1})$, i.e. we split each
mixture component $\varphi_j$ into $2D+1$ components
$\delta_i\varphi_{ji}$, $i = 0, \dotsc, 2D$, and propagate them
forward using the Unscented Transform. For notational convenience, we
define this operation on a Gaussian mixture as
$\mathcal{F}_{n}(f,p(x_{n-1}))$, such that
\begin{align}
  \tildew{p}(x_{n}) &=\mathcal{F}_{n}(f,p(x_{n-1})) \!=\! \int p(x_n|x_{n-1} )p(x_{n-1}) \mathrm{d} x_{n-1}\nonumber\\
  &=\sum_{j=0}^{M-1} 
  \sum_{i=0}^{2D}\beta_{j} {\delta}_{i} \int p(x_n|x_{n-1} ){\varphi}_{ij}(x_{n-1}) \mathrm{d} x_{n-1}\nonumber \\
  &\equiv \GSUMtilde{l}{M(2D+1)-1}{\gamma}{(x_{n})},
\label{equ:GMMap}
\end{align}
where $\gamma_{l}=\delta_{i}\beta_{j}$. We compute the moments of the
mixture components $\varphi_l$ by means of the Unscented Transform.


\subsection{Mixture Reduction}
\label{subsec:mxtrred}
Up to this point, we have considered the case where a density with
known mean and variance has been represented by a GMM, which could
subsequently be used to estimate the predicted state distribution.
Incorporating these steps into an recursive state estimator for time
series, there is an exponential growth in the number of mixture
components in~\eqref{equ:GMMap}.  One way to mitigate this effect is
to represent the estimated densities by a mixture model with a fixed
number of components \cite{Anderson2005}. To keep the number of
mixture components constant we can reduce them at each time step
\cite{Anderson2005}. 

A straightforward and fast approach is to drop the Gaussian components
with the lowest weights. Such omissions, however, can result in poor
performance of the filter
\cite{kitagawa1994two}. Kitagawa~\cite{kitagawa1994two} suggested to
repeatedly merge a pair Gaussian components. A pair is selected with
lowest distance in terms of some distance metric.
We evaluated multiple distance metrics, e.g. the $L_2$ distance
\cite{williams2003cost}, the KL divergence
\cite{runnalls2007kullback}, and the Cauchy Schwarz divergence
\cite{kampa2011closed}. In this paper, we used the symmetric KL
divergence \cite{kitagawa1994two}, $ D(p,q)= ( \mathcal{KL}\,(p|q) +
\mathcal{KL}\,(q|p))/2 $, which outperformed aforementioned distance
measures for mixture reduction in filtering.  


\subsection{Filtering}
\label{sec:Gsum}
In the following, subsume all derivations in our multi-modal
non-linear state estimator, whose time and measurement updates are
summarised in the following.

\subsubsection{Time Update}
\label{subsec:tmeupdate}
Assume that the filter distribution $p(x_{n-1|n-1})$ is represented by
a GMM with $M$ components.  The time update, i.e. the one-step ahead
predictive distribution is given by
\begin{equation}
\label{equ:predgeneric}
\begin{aligned}
p(x_{n|n-1})&=\int p(x_{n}|x_{n-1})p(x_{n-1|n-1})\, \mathrm{d} x_{n-1}. \\
\end{aligned}
\end{equation}
This integral can be evaluated as $\mathcal{F}_{n}(f,p(x_{n-1|n-1}))$,
such that we obtain a GMM representation of the time update
\begin{align}
\label{equ:timeupdate}
p(x_{n|n-1})&=\GSUMtilde{j}{M(2D+1)-1}{\gamma}{(x_{n|n-1})},
\end{align} 
as detailed in~\eqref{equ:GMMap}.
\subsubsection{Measurement Update}
\label{subsec:Measurementupdte}
The measurement update can be approximated up to a normalisation constant by 
\begin{equation}
\label{equ:msrupdateapprox}
p(x_{n|n}) \propto p(y_{n}|x_{n}) p(x_{n|n-1}),
\end{equation}
where $p(x_{n|n-1})$ is the time update \eqref{equ:timeupdate}. We
now apply a similar operation as in \eqref{equ:GMMap} with $h$ as
non-linear function and obtain
\begin{equation}
\label{equ:nonlnrmeasr}
p(y_{n|n-1}) = \mathcal{F}_{n}(h,p(x_{n|n-1})).
\end{equation}
Substituting \eqref{equ:nonlnrmeasr} and \eqref{equ:timeupdate} in
\eqref{equ:msrupdateapprox} yields the measurement update, i.e. the
filtered state distribution
\begin{align}
p(x_{n|n})&\propto \GSUM{i}{2D}{\delta}{(y_{n|n-1})}   \GSUMtilde{j}{M(2D+1)-1}{\gamma}{(x_{n|n-1})}\nonumber \\
   &\equiv \GSUM{l}{M(2D+1)^2-1}{\beta}{(x_{n|n})}.
\label{equ:fwdfilter}
\end{align}
We calculate the measurement update for each pair $\varphi_{i}$
and $\varphi_{j}$.  Recalling that $\varphi_{l} (x_{n|n}) = \N (x |
\mu^{ij}_{n|n},\Sigma^{ij}_{n|n} )$ for $i = 0,\dotsc,2D$ and $j =
0,\dotsc,(2D+1)^2-1$, the measurement updates \cite{sarkka2008unscented}
and weight updates (Gaussian Sum \cite{alspach1972nonlinear}) can be derived by
\begin{equation}
\begin{aligned}
  K_{n}^{j}&= \tildew{\Gamma}_{n|n-1}^{j} \big(
  \tildew{\Sigma}_{n|n-1}^{j} \big)^{-1}, \\
  \mu_{n|n}^{ij}&= \mu_{n|n-1}^{i}+ K_{n}^{j}\big(  y- \tildew{\mu}_{n|n-1}^{j} \big),\\
  \Sigma_{n|n}^{ij}&= \Sigma_{n|n-1}^{i} - K_{n}^{j}\big(
  \tildew{\Sigma}_{n|n-1}^{j} \big) {K_{n}^{j}}^{T},\\
  \beta_{i,j}&= \frac{\delta_{i}\,\gamma_{j}\N(x=y\,|\,
    \tildew{\mu}_{n|n-1}^{j},\tildew{\Sigma}_{n|n-1}^{j})}{\displaystyle
    \sum\nolimits_{k,l}
    \delta_{l}\,\gamma_{k}\N(x=y\,|\,\,\tildew{\mu}_{n|n-1}^{k},\tildew{\Sigma}_{n|n-1}^{k})}
  ,
\end{aligned}
\end{equation}
where $\tildew{\Gamma}_{n|n-1}^{j}$ is the cross covariance matrix
$\mathrm{cov}(x_{n-1}, x_n)$ determined via the Unscented Transform
\cite{sarkka2008unscented}. 

After the measurement update, we reduce the \mbox{$M(2D+1)^2$} mixture
components in the GMM, see~\eqref{equ:fwdfilter}, to $M$ according to
Section~\ref{subsec:mxtrred}.
\section{RESULTS}
\label{sec:results}
We evaluated our proposed filtering algorithm on data generated from
standard one-dimensional non-linear dynamical systems. Both the UKF
and the Multi-Modal Filter (M-MF) use the same parameters for the
Unscented Transform, i.e. $\alpha=1$, $\beta=2$ and $\kappa=2$. The
prior $p(x_{0})$ is a standard Gaussian $p(x_{0})=
\N(x|0,1)$. Densities in the M-MF are represented by a Gaussian
mixture with $M=3$
components. 
The mean of the filtered state is estimated by the first moment of
this Gaussian mixture. The employed Particle filter (PF) is a standard
particle filter with residual re-sampling scheme. The PF-UKF
\cite{van2000unscented} on the other hand uses the Unscented Transform
as proposal distribution and the UKF for filtering. Unlike our M-MF
the PF-UKF uses random sampling and the UKF. We use the root mean
square error (RMSE) and the predictive Negative Log-Likelihood (NLL)
per observation as metrics to compare the performance of the different
filters. Lower values indicate better performance. The results in
Table \ref{tbl:results} were obtained from 100 independent simulations
with $T=100$ time steps for each of the following system models.

\subsection{Non-stationary Time Series}
\label{subsec:Non-stationary}
We tested the different filters on a standard system, the Uniform
Non-stationary Growth Model (UNGM) from
\cite{doucet2001sequential} 
\begin{align}
\hspace{-2mm}x_{n} &\!=\! \tfrac {x_{n-1}}{2}  + \tfrac{25x_{n-1}}{1+x^{2}_{n-1}} + 8\cos(1.2(n-1))\!+\!w, \label{eq:ungm}\\ 
y_{n} &\!=\! \tfrac{x^{2}_{n}}{20} + v,\nonumber
\end{align}
where $ w\sim\mathcal{N}(0,1), v\sim\mathcal{N}(0,1)$.  The true state
density of the non-stationary model stated above alternated between
multi-modal and uni-modal distributions. The switch from uni-modal to
a bi-modal density occurred when the mean was close to zero. The
quadratic measurement function makes it difficult to distinguish
between the two modes as they are symmetric around zero.  This
symmetry posed a substantial challenge for several filtering
algorithms. The PF lost track of the state dynamics as $N=500$
particles failed to capture true density especially in its tails,
which led to degeneracy~\cite{doucet2001sequential}. The particle
filters in Table \ref{tbl:results} are in their standard form and
performance may improve if advanced techniques are used
\cite{doucet2001sequential}, as can be seen from Gibbs
filter~\cite{deisenroth2011general}. The proposed M-MF could track
both modes and, hence, led to more consistent estimates. The RMSE
performance of both multi-modal filter (M-MF) and UKF are similar with
M-MF performing better in terms of a lower mean error and standard
deviation. The main advantage of the M-MF is its ability to capture
the uncertainty. The NLL values of the M-MF were significantly better
than the UKF even when the same parameters are used to calculate the
Unscented Transform, see Table~\ref{tbl:results}.

We tested the UNGM with an alternative measurement function
$h(x)=5\sin(x)$. For this function, the performance of the EKF is best
in terms of RMSE, since its estimates are more stable. The proposed
M-MF could track multiple modes, which resulted in significantly
better performance in terms of NLL. Moreover, the filter performance
was consistently stable, indicated by the small standard deviation
values for the NLL measure.
\begin{table}[!t]
\begin{center}
\resizebox{\hsize}{!} {
\begin{tabular}{c||c|c||c|c|c|c}
 \rule[-1ex]{0pt}{2.5ex}  & \multicolumn{2}{|c||}{Stationary}&\multicolumn{4}{|c}{Non-Stationary} \\ 
 
& \multicolumn{2}{|c||}{$h(x)=5\sin(x)$}&\multicolumn{2}{|c|}{$ h(x)=x^{2}/20$}&\multicolumn{2}{|c}{$h(x)=5\sin(x)$} \\ 
 \hline
 \rule[-1ex]{0pt}{2.5ex}  & RMSE & NLL & RMSE & NLL  & RMSE & NLL  \\  
 \hline\hline
 \rule[-1ex]{0pt}{2.5ex} EKF & 7.5 $\pm$ 0.4 & 340.7 $\pm$ 38.6 & 10.9 $\pm$ 1.3 & 103.6 $\pm$ 29.2 & 10.4 $\pm$ 0.4 & 662.2 $\pm$ 98.5 \\ 
 \hline 
 \rule[-1ex]{0pt}{2.5ex} UKF & 12.2 $\pm$ 2.3 & 64.2 $\pm$ 46.7 & 6.5 $\pm$ 1.9 & 15.1 $\pm$ 12.3   & 10.0 $\pm$ 24  & 42.0 $\pm$ 32.8\\ 
 \hline 
 \rule[-1ex]{0pt}{2.5ex} M-MF & \textbf{1.4 $\pm$ 0.4} & \textbf{1.0 $\pm$ 0.1} & {6.1 $\pm$ 1.2} & \textbf{1.7 $\pm$ 0.6} & 9.4 $\pm$ 2.7 & {3.7 $\pm$ 1.3} \\ 
 \hline
 \rule[-1ex]{0pt}{2.5ex} PF & {2.4 $\pm$ 0.03} & N/A & {8.3 $\pm$ 6.1} & N/A & 11.2 $\pm$ 3.6 & N/A \\ 
 \hline
 \rule[-1ex]{0pt}{2.5ex} PF-UKF & {16.8 $\pm$ 0.1} & N/A & {13.1 $\pm$ 1.8} & N/A & 20.0 $\pm$ 2.4 & N/A \\ 
 \hline
  \rule[-1ex]{0pt}{2.5ex} Gibbs Filter & {3.62 $\pm$ 0.4} & {3.06 $\pm$ 0.03} & \textbf{{3.97 $\pm$ 0.4}} & {2.18 $\pm$ 0.1} & \textbf{8.64 $\pm$ 0.4} & \textbf{{3.57 $\pm$ 0.08}}    
\end{tabular}}
\caption{Average performances of the filters are shown along with
  standard deviation. Lower values are better. The M-MF filter
  performs better than the standard filtering methods.}
\label{tbl:results}
\end{center}
\vspace{-0.8mm}
\end{table}
\subsection{Stationary Time Series}
\label{subsec:stationary}
We modified the UNGM described above to 
$$x_{n}= \tfrac {x_{n-1}}{2} +
\tfrac{25x_{n-1}}{1+x^{2}_{n-1}} + w\,,\quad w\sim\mathcal{N}(0,1)\,,$$
\vspace{-0.1mm}
by dropping the time dependant cosine term from~\eqref{eq:ungm} and
use a sinusoidal measurement function
$$y_{n}=5\sin(x_{n}) + v\,,\quad v\sim\mathcal{N}(0,1).$$ 
We see from the results in Table~\ref{tbl:results} that the UKF was
outperformed by all other deterministic filters. The failures of the
UKF and the PF-UKF are attributed to their overconfident predictions
for sinusoidal functions, which confirms the results in
\cite{deisenroth2012robust}. The EKF approximates these sinusoidal
functions better but both the UKF and EKF fail to capture the
multi-modal nature of system
dynamics 
Thus, the EKF and UKF are inconsistent for the model and settings used
in this experiment. The proposed M-MF on the other hand performed
consistently better in terms of RMSE and NLL values. Moreover, the
small standard deviation of the NLL suggests that our proposed M-MF is
consistent and stable.

%



\section{CONCLUSION AND FUTURE WORK}
\label{sec:Conclusion}
In this paper, we presented the M-MF, a Gaussian mixture based
multi-modal filter for state estimation in non-linear dynamical
systems. Multi-modal densities are represented by Gaussian mixtures,
whose parameters are computed in closed form.  We demonstrated that
the M-MF achieves superior performance compared to state-of-the-art
state estimators and consistently captures the uncertainty in
multi-modal densities.

In future work, we will evaluate the significance of the scaling
parameter $\alpha$ and its impact on higher moments of the
approximations. The effect of the mixture reduction techniques also
will be investigated to achieve better filter
performance. Moreover, we will extend the M-MF to a forward-backward
smoothing algorithm.

\bibliographystyle{IEEEbib}
\bibliography{refs.bib}

\end{document}